\documentclass[a4paper]{article}
\usepackage[T1]{fontenc}
\usepackage[utf8]{inputenc}

\usepackage[backend=bibtex,sortcites]{biblatex}

\bibliography{introduction_SMT_for_computer_algebra}

\usepackage{dmbiblatex,dmbiblatexlikelncs,ifthen,amsfonts,amssymb,amsmath,tikz,proof,paralist,array,listings,amsthm}
\usepackage[pdfauthor={David Monniaux},pdftitle={A survey of satisfiability modulo theory}]{hyperref}
\newtheorem{example}{Example}

\title{A Survey of Satisfiability Modulo Theory%
\thanks{The research leading to these results has received funding from the
\href{http://erc.europa.int/}{European Research Council} under the European
Union's Seventh Framework Programme (FP/2007-2013) / ERC Grant Agreement
nr.~306595 \href{http://stator.imag.fr/}{\mbox{``STATOR''}}.}}

\author{David Monniaux\protect\\
{\footnotesize Univ. Grenoble Alpes, VERIMAG, F-38000 Grenoble, France}\protect\\%
{\footnotesize CNRS, VERIMAG, F-38000 Grenoble, France}}

\newboolean{fullversion}

\lstdefinelanguage{smtlib}[]{Lisp}{morekeywords={set-option,set-logic,declare-fun,check-sat,get-model,model,define-fun,unsat,sat}}
\newcommand{\true}{\mathbf{t}}
\newcommand{\false}{\mathbf{f}}

\newcommand{\abstr}[1]{#1^\sharp}

\newcommand{\lneg}[1]{\bar{#1}}

\newcommand{\ZZ}{\mathbb{Z}}
\newcommand{\RR}{\mathbb{R}}

\newcommand{\defn}{\stackrel{\triangle}{=}}
\newcommand{\ve}[1]{\vec{#1}}
\newcommand{\dotp}[2]{{#1} \cdot {#2}}
\newcommand{\update}[3]{\mathit{update}(#1,#2,#3)}
\newcommand{\ite}[3]{\mathit{ite}(#1,#2,#3)}

\newcommand{\soft}[1]{\textsc{#1}}

\newcommand{\formularef}[1]{(\ref{#1})}

\makeatletter
\g@addto@macro\normalsize{%
\abovedisplayskip 5.0\p@ \@plus2\p@ \@minus4\p@
\abovedisplayshortskip \z@ \@plus2\p@
\belowdisplayshortskip 3\p@ \@plus2\p@ \@minus2\p@
\belowdisplayskip \abovedisplayskip
}
\g@addto@macro\small{%
\abovedisplayskip 7.5\p@ \@plus2\p@ \@minus4\p@
\abovedisplayshortskip \z@ \@plus2\p@
\belowdisplayshortskip 3\p@ \@plus2\p@ \@minus2\p@
\belowdisplayskip \abovedisplayskip
}
\renewcommand{\paragraph}{%
  \@startsection{paragraph}{4}%
  {\z@}{0.8ex \@plus 0ex \@minus 1ex}{-1em}%
  {\normalfont\normalsize\bfseries}%
}

\def\@IEEEsectpunct{.\ \,}

\newlength{\sectionspace}
\renewcommand\section{\@startsection{section}{1}{\z@}%
                       {-18\sectionspace \@plus -4\sectionspace \@minus -4\sectionspace}%
                       {12\sectionspace \@plus 4\sectionspace \@minus 4\sectionspace}%
                       {\normalfont\large\bfseries\boldmath
                        \rightskip=\z@ \@plus 8em\pretolerance=10000 }}

\makeatother

\begin{document}
\maketitle

\begin{abstract}
Satisfiability modulo theory (SMT) consists in testing the satisfiability of first-order formulas over linear integer or real arithmetic, or other theories. In this survey, we explain the combination of propositional satisfiability and decision procedures for conjunctions known as DPLL(T), and the alternative ``natural domain'' approaches. We also cover quantifiers, Craig interpolants, polynomial arithmetic, and how SMT solvers are used in automated software analysis.
\end{abstract}

\section{Introduction}
\emph{Satisfiability modulo theory} (SMT) solving consists in deciding the satisfiability of a first-order formula with unknowns and relations lying in certain theories. For instance, the following formula has no solution $x,y \in \RR$:%
\footnote{This survey focuses on linear and polynomial numeric constraints over integers and reals.
SMT however encompasses theories as diverse as character strings, inductive data structures, bit-vector arithmetic, and ordinary differential equations.} 
\begin{equation}
(x \leq 0 \lor x + y \leq 0) \land y \geq 1 \land x \geq 1 \,.
\end{equation}
The formula may contain negations ($\neg$), conjunctions ($\land$), disjunctions ($\lor$) and, possibly, quantifiers ($\exists$, $\forall$).

A \emph{SMT-solver} reports whether a formula is satisfiable, and if so, may provide a \emph{model} of this satisfaction; for instance, if one omits $x \geq 1$ in the preceding formula, then its solutions include $(x=0,y=1)$.
Other possible features include dynamic addition and retraction of constraints, production of proofs and Craig interpolants (Sec.~\ref{sec:interpolants}), and optimization (Sec.~\ref{sec:optimization}). 
SMT-solving has major applications in the formal verification of hardware, software, and control systems.

Quantifier-free SMT subsumes Boolean satisfiability (SAT), the canonical NP-complete problem, and certain classes of formulas accepted by SMT-solvers belong to higher complexity classes or are even undecidable.
This has not deterred researchers from looking for algorithms that, in practice, solve many relevant instances at reasonable costs.
Care is taken that the worst-case cost does not extend to situations that can be dealt with more cheaply.

\ifthenelse{\boolean{fullversion}}{%
The core idea underlying all successful approaches is \emph{learning} during search: if one branch of the search space is unfruitful, the system learns a lemma explaining why --- for instance, from trying unsuccessfully with $x=1$ and $y=2$, it may learn a lemma that there is no solution with $x \geq 0 \land y \geq 1$.
Hopefully this lemma is quite general and kills not just one particular unsuccessful attempt at a solution, but many similar ones.}{}

Most SMT solvers follow the DPLL(T) framework (Sec.~\ref{sec:DPLLT}):
a CDCL\ifthenelse{\boolean{fullversion}}{\footnote{\emph{Constraint-driven clause learning} is an evolution of the DPLL approach to SAT-solving, thus the name.}}{}
solver for SAT (Sec.~\ref{sec:CDCL}) is used to traverse the Boolean structure, and conjunctions of atoms from the formula are passed to a solver for the theory.
This approach limits the interaction between theory values and Boolean reasoning, which led to the introduction of \emph{natural domain} approaches (Sec.~\ref{sec:natural_domain}).
Finally, we shall see in Sec.~\ref{sec:beyond} how to go beyond mere quantifier-free satisfiability testing, by handling quantifiers, providing Craig interpolants, or providing optimal solutions.
Let us now first see a few generalities, and how SMT-solving is used in practice.

\subsection{Generalities}
Consider quantifier-free propositional formulas, that is, formulas constructed from \emph{unknowns} (or \emph{variables}) taking the values ``true'' ($\true$) and ``false'' ($\false$) and propositional connectives $\lor$ (or), $\land$ (and), $\neg$ (not); $\lneg x$ shall be short-hand for $\neg x$.%
\footnote{Further propositional connectives, such as exclusive-or, or ``let $x$ be $e_1$ in $e_2$'' constructs may be also considered.}
A formula is: in \emph{negation normal form} (NNF) if the only $\neg$ connectives are at the leaves of its syntax tree (that is, wrap around unknowns but not larger formulas); a \emph{clause} if it is a disjunction of literals (a literal is an unknown or its negation); in \emph{disjunctive normal form} (DNF) if it is a disjunction of conjunctions of literals; in \emph{conjunctive normal form} (CNF) if it is a conjunction of clauses.
If $A$ implies $B$, then $A$ is \emph{stronger} than $B$ and $B$ \emph{weaker} than $A$.
Uppercase letters ($F$) shall denote formulas, lowercase letters ($x$) unknowns, and lowercase bold letters ($\ve{x}$) vectors of unknowns.

\emph{Satisfiability testing} consists in deciding whether there exists a \emph{satisfying assignment} (or \emph{solution}) for these unknowns, that is, an assignment making the formula true. For instance, $a=\true, b=\true, c=\false$ is a satisfying assignment for
$(a \lor c) \land (b \lor c) \land (\lneg a \lor \lneg c)$.
In case the formula is satisfiable, a solver is generally expected to provide such a satisfying assignment; in the case it is unsatisfiable, it may be queried for an \emph{unsatisfiable core}, a subset of the conjunction given as input to the solver that is still unsatisfiable.

\ifthenelse{\boolean{fullversion}}{%
Propositional satisfiability (SAT) is the canonical NP-complete problem, since the introduction of the concept of NP-completeness by Cook \cite{Cook:1971:CTP:800157.805047}, Karp \cite{Karp1972} and Levin \cite{Levin73,Trakhtenbrot84}. It remains NP-complete even if the formulas are in conjunctive normal form (CNF-SAT), even if the clauses are restricted to have only 3 literals (3CNF-SAT).}

\emph{Satisfiability modulo theory} extends propositional satisfiability by having some atomic propositions be predicates from a theory. For instance, $(x>0 \lor c) \land (y>0 \lor c) \land (x\leq 0 \lor \lneg c)$ is a formula over linear rational arithmetic (LRA) or linear integer arithmetic (LIA), depending on whether $x$ and $y$ are to be interpreted over the rationals or integers.

Different unknowns may range in different sets; for instance
$f(x)\neq f(y) \land x=z+1 \land z=y-1$ has unknowns $f: \ZZ \rightarrow \ZZ$ and $x,y,z \in \ZZ$. This formula is said to be over the combination of \emph{uninterpreted functions and linear integer arithmetic} (UFLIA).
In this formula, $f$ is said to be uninterpreted because we give no definition for it; we shall see in Sec.~\ref{sec:UF} that this formula has no satisfying assignment and how to establish this fact automatically.

\ifthenelse{\boolean{fullversion}}{
The language of formulas may be enriched by allowing quantifiers $\exists$ and $\forall$. A variable is is \emph{free} if it is not introduced by a quantifier, \emph{bound} otherwise; a formula with no free variable is \emph{closed}.
For instance, $\forall x~ x \geq y \Rightarrow x \geq 1$, over linear rational arithmetic, is equivalent to $y \geq 1$ (in the sense that both have the same satisfying assignments over their free variable~$y$).
A closed formula is equivalent to $\true$ or $\false$.

All quantifiers in a formula can be moved to the front, if necessary by renaming some variables; such a formula is said to be \emph{prenex}.
For instance, $\forall x~ x > 1 \lor \exists y~ x \leq y \leq 1$ may be converted to $\forall x \exists y~ x > 1 \lor x \leq y \leq 1$. The \emph{quantifier prefix} is then $\forall x \exists y$, summarized $\forall \exists$.

Converting a formula with quantifiers into an equivalent formula without quantifiers is known as \emph{quantifier elimination}. It is possible to decide the satisfiability of formulas containing quantifiers by first eliminating these quantifiers and then testing for satisfiability (or even by eliminating the implicit existential quantifier in front of all free variables of the formula), but this is not the only way.

The addition of quantifiers tends to make satisfiability testing more difficult. A closed formula in prenex form has a quantifier prefix of the form $\exists^* \forall^* \exists^* \dots$ or $\forall^* \exists^* \forall^* \dots$, where $\forall^*$ (resp. $\exists^*$) denotes zero or more occurrences of $\forall$ (resp. $\exists$). $\exists^*$ (resp. $\forall^*$) denotes a \emph{block} of existential (resp. universal) quantifiers.
Deciding the truth of closed $\exists^*$ propositional formulas is thus propositional satisfiability (SAT), NP-complete; thus deciding the truth of closed $\forall^*$ formulas is co-NP-complete.
These results extend to the \emph{polynomial hierarchy}: deciding the truth of $\exists^* \forall^* \dots$ (resp. $\forall^* \exists^* \dots$) with $n$ alternating quantifier blocks is $\Sigma^p_n$-complete (resp. $\Pi^p_n$-complete).
Deciding the truth of propositional formulas with arbitrary alternation of quantifiers, a problem known as true quantified Boolean formula (TQBF or QBF), is PSPACE-complete.

Adding quantifiers may quickly make the class of formulas under consideration \emph{undecidable}, that is, there is no algorithm for deciding their satisfiability. For instance, the combination of uninterpreted functions and linear integer arithmetic is decidable for quantifier-free formulas, but adding universal quantification over the integers yields an undecidable class.%
\footnote{In short, given a Turing machine $M$ it is possible to constrain the values of an uninterpreted function $f: \ZZ^2 \rightarrow \ZZ$ so that $f(i,j)$ is the $j$-th symbol of the tape of $M$ after $i$ steps.}
}{}

\subsection{The SMT-LIB Format and Available Theories}
\label{sec:SMT-LIB}

SMT solvers can be used
\begin{inparaenum}[i)]
\item as a library, from an application programming interface, typically from C/C++, Java, Python, or OCaml
\item as an independent process, from a textual representation, possibly through a bidirectional pipe.
\end{inparaenum}

APIs for SMT-solvers are not standardized, though there have been efforts such as JavaSMT%
\footnote{\url{https://github.com/sosy-lab/java-smt} \cite{Karpenkov_et_al_VSTTE2016}}
to provide a common layer for several solvers.
In contrast, much effort has been put into designing and supporting the common SMT-LIB \cite{BarFT-SMTLIB} format, a textual representation (Listing~\ref{lst:example_uf});
some solvers support other languages than SMT-LIB, sometimes alongside it.
Libraries of benchmark problems, sorted according to the theories involved and the presence or absence of quantifiers (Tab.~\ref{tab:SMT-LIB}), are available in that format.
New theories are proposed; for instance, a theory for constraints over IEEE-754 floating-point arithmetic \cite{IEEE-754} is under evaluation.

\begin{lstlisting}[language=smtlib,float,caption={Example of SMT-LIB 2 file. Assertions $x \geq 0$, $y \leq 0$, $f(x) \neq f(y)$ and $x + y \leq 0$ are added, then the problem is checked to be unsatisfiable. The last assertion is retracted and replaced by $x + y \leq 1$, the problem becomes satisfiable and a model is requested (see Listing~\ref{lst:example_uf_answer})},label=lst:example_uf]
(set-logic QF_UFLIA)
(set-option :produce-models true)
(declare-fun x () Int)
(declare-fun y () Int)
(declare-fun f (Int) Int)
(assert (>= x 0))
(assert (>= y 0))
(assert (distinct (f x) (f y)))
(push 1)
(assert (<= (+ x y) 0))
(check-sat)
(pop 1)
(assert (<= (+ x y) 1))
(check-sat)
(get-model)
\end{lstlisting}
\begin{lstlisting}[language=smtlib,float,caption={Z3's answers to the SMT-LIB Listing~\ref{lst:example_uf}},label=lst:example_uf_answer]
unsat
sat
(model 
  (define-fun y () Int 1)
  (define-fun x () Int 0)
  (define-fun f ((x!1 Int)) Int
    (ite (= x!1 0) 2
    (ite (= x!1 1) 3
      2)))
)
\end{lstlisting}

\begin{table}[tb]
\begin{center}
\begin{tabular}{l|>{\tt}r}
linear real arithmetic & LRA\\
linear integer arithmetic & LIA\\
linear mixed integer and real arithmetic & LIRA\\
bit-vector arithmetic & BV\\
nonlinear (polynomial) real arithmetic & NRA\\
nonlinear (polynomial) integer arithmetic & NIA\\
nonlinear (polynomial) mixed integer and real arithmetic & NIRA\\
\hline
uninterpreted functions & UF \\
arrays & A / AX\\
\hline
quantifier-free & QF\_
\end{tabular}
\end{center}

\caption{Categories of formulas in SMT-LIB; e.g. \hbox{\texttt{QF\_UFLIA}} means quantifier-free combination of uninterpreted functions.}
\label{tab:SMT-LIB}
\end{table}

\ifthenelse{\boolean{fullversion}}{%
A major difference between SMT-LIB~1 and 2 is that in SMT-LIB~1, a satisfiability problem could be defined, but the language neither contained commands such as ``retract the last two constraints'' or ``check and return a model if satisfiable'', neither specified how values were to be returned by the solver to the calling application, thus forcing the use of an API for many purposes, while in SMT-LIB~2 all these functions may be used through a bidirectional pipe (see Listings~\ref{lst:example_uf} and \ref{lst:example_uf_answer} for examples).}{}

Alas, some features, such as quantifier elimination or the extraction of Craig interpolants (Sec.~\ref{sec:interpolants}) do not have standard commands.
Furthermore, not all tools implement all operators and commands following the standard.

\subsection{Use in Program Analysis Applications}
\label{sec:program_analysis}
A major use of SMT-solvers is the analysis of software.
In most cases (but not always), the solutions of the formula to be tested for satisfiability correspond to execution traces of the software verifying certain desirable or undesirable properties: for instance traces going into error states.

\subsubsection{Symbolic Execution}
In \emph{symbolic program execution} \cite{King:1976:SEP:360248.360252}, a program is executed as though operating on symbolic inputs.
Along a straight path in the program, the semantics of the instructions and tests encountered accumulate as a \emph{path condition}, expressing the relationship between the final values and the inputs.
In case a branching instruction is encountered, the analyzer tests whether either branch may be taken by checking for a solution to the conjunction of the path condition and the guard associated with the branch: branches for which a solution is known not to exist are not retained for the rest of the analysis.
The analysis thus explores a tree of possible executions, which in general does not cover all possible executions of the program: this is acceptable in bug-finding applications.

Pure symbolic execution may prove infeasible due to the large number of paths to explore. This is especially true if the program involves loads and writes to memory, due to the \emph{aliasing} conditions to test (``does this read correspond to this write?'').
Because of this, often what is done is a mixture of concrete and symbolic execution, dubbed \emph{concolic}: sometimes a non-symbolic value is picked (e.g. memory allocation addresses) for simpler execution.
In \emph{whitebox fuzzing}, concolic execution is applied from symbolic values coming from external inputs (files, network communications) so as to reach security hazards~\cite{Godefroid:2012:SWF:2090147.2094081}.

\subsubsection{Inductiveness Check and Bounded Model Checking}
In some other cases \cite{Henry_Monniaux_Moy_SAS2012,Henry_Monniaux_Moy_TAPAS2012,Gawlitza_Monniaux_LMCS12}, the formula encodes the full set of executions between two control locations in a program, such that there is no looping construct between these locations: one Boolean variable is added per control location, expressing whether or not the execution goes through that location.

In the Floyd-Hoare approach to proving the correctness of programs (see e.g. \cite{Winskel_1993}), the user is prompted for an inductive invariant for each looping construct: a formula $I$ that holds at loop initiation, and that, if it holds at one loop iteration, holds at the next (\emph{inductiveness}). In other words, there is no execution of the loop guard and loop body that starts in $I$ and ends in~$\neg I'$ ($I'$ is $I$ where the variables are renamed in order to express their final, not initial, values). In modern tools,
the loop guard and body are turned into a first-order formula that is conjoined with $I$ and $\neg I'$, then checked for unsatisfiability;
or equivalently through a \emph{weakest precondition} computation, as in Frama-C \cite{DBLP:conf/sefm/CuoqKKPSY12}.

\begin{example}
Consider the array fill program (assume $n \geq 0$):
\begin{lstlisting}[language=C,firstline=3,lastline=4]
  int t[n];
  for(int i=0; i<n; i++) t[i] = 42;
\end{lstlisting}
In order to prove the postcondition $\forall k~ 0 \leq k < \lstinline|n| \Rightarrow \lstinline|t|[k]=42$, one needs the loop invariant
\begin{equation}
I \defn (0 \leq i \leq n) \land (\forall k~ 0 \leq 0 \leq k < i \Rightarrow t[k]=42) \,.
\end{equation}

\noindent The inductiveness condition is
\begin{equation}
(I \land i < n) \Rightarrow
I[i \mapsto i+1, t \mapsto \update{t}{i}{42}] \,,
\end{equation}
where $\update{t}{i}{42}$ is the array $t$ where $i$ has been replaced by $42$, and $I[i \mapsto x]$ is formula $I$ where $i$ has been replaced by~$x$. 
This condition is checked by showing that the negation of this formula is unsatisfiable --- after Skolemization:
\begin{multline}
(0 \leq i \leq n) \land (\forall k~ 0 \leq k < i \Rightarrow t[k]=42)
\land i < n \\ \land
\left(\neg (0 \leq i+1 \leq n) \lor
(0 \leq k_0 \leq i \land \update{t}{i}{42}[k_0] \neq 42)\right) \,.
\end{multline}
$\update{t}{i}{42}[k_0]$ expands into $\ite{k_0=i}{42}{t[k_0]}$ where $\ite{a}{b}{c}$ means ``if $a$ then $b$ else $c$''.
The universal quantifier is instantiated with $k=k_0$, a new unknown $t_k = t[k]$ is introduced to handle the \emph{uninterpreted function} $f$ (Sec.~\ref{sec:UF}) and the resulting problem is solved over linear integer arithmetic (Sec.\ref{sec:LIA}). 
\end{example}

\section{The DPLL(T) Architecture}
Most SMT-solvers follow the DPLL(T) architecture: a solver for pure propositional formulas, following the DPLL or CDCL class of algorithms, drives decision procedures for each theory (e.g. linear arithmetic) by adding or retracting constraints and querying for satisfiability.
\emph{DPLL(T) and decision procedures for many interesting logics are explained in more detail in e.g. \cite{Kroening_Strichman_08,BradleyManna07}.}

\subsection{CDCL Satisfiability Testing}
\label{sec:CDCL}
\emph{We shall only give a cursory view of satisfiability testing and refer the reader to e.g. \cite{HandbookOfSAT2009} for more in-depth treatment.}

Many algorithms for satisfiability testing for quantifier-free formulas only accept formulas in conjunctive normal form (conjunction of clauses). Naive conversion into conjunctive normal form, by application of distributivity of $\lor$ over $\land$, incurs an exponential blowup.
It is however possible to construct, from any formula $F$, a formula $F'$ in CNF but with additional free variables, such that any satisfying assignment to $F$ can be extended to a satisfying assignment on $F'$ and any satisfying assignment on $F'$, restricted to the free variables of $F$, is a satisfying assignment of~$F$.
\emph{Tseitin's encoding} is the simplest way to do so: to any subformula $e_1 \land e_2$ of $F$, associate a new propositional variable $x_{e_1 \land e_2}$ and constrain it such that it is equivalent to $e_1 \land e_2$ by clauses $\neg x_{e_1 \land e_2} \lor e_1$, $\neg x_{e_1 \land e_2} \lor e_2$, $\neg e_1 \lor \neg e_2 \lor x_{e_1 \land e_2}$ (and similarly for $e_1 \lor e_2$).

\begin{example}
Consider
\begin{equation}\label{formula:before_tseitin}
\left( (a \land \lneg{b} \land \lneg{c})
  \lor (b \land c \land \lneg{d}) \right) \land
(\lneg{b} \lor \lneg{c}) \,.
\end{equation}
Assign propositional variables to sub-formulas:
\begin{equation}
\begin{array}{c@{\hskip 2em}c@{\hskip 2em}c@{\hskip 2em}c@{\hskip 2em}c}
e \equiv a \land \lneg{b} \land \lneg{c} &
f \equiv b \land c \land \lneg{d} &
g \equiv e \lor f &
h \equiv \lneg{b} \lor \lneg{c} &
\phi \equiv g \land h \,;
\end{array}
\end{equation}
these equivalences are turned into clauses:
\begin{equation}\label{system:after_tseitin}
\begin{array}{c@{\hskip 2em}c@{\hskip 2em}c@{\hskip 2em}c}
\lneg{e} \lor a & \lneg{e} \lor \lneg{b} & \lneg{e} \lor \lneg{c}
  & \lneg{a} \lor b \lor c \lor e\\
\lneg{f} \lor b & \lneg{f} \lor c & \lneg{f} \lor d
  & \lneg{b} \lor \lneg{c} \lor d \lor f\\
\lneg{e} \lor g & \lneg{f} \lor g &
  \lneg{g} \lor e \lor f \\
b \lor h & c \lor h & \lneg{h} \lor \lneg{b} \lor \lneg{c}\\
\lneg{\phi} \lor g & \lneg{\phi} \lor h & \lneg{g} \lor \lneg{h} \lor \phi & \phi \,.
\end{array}
\end{equation}

The model $(a,b,c,d) = (\true, \false, \false, \true)$ of \formularef{formula:before_tseitin} is extended by $(e,f,g) = (\true, \false, \true)$, producing a model of the system of clauses \formularef{system:after_tseitin}, i.e., the conjunction of these clauses. Conversely, any model of that system, projected over $(a,b,c,d)$, yields a model of \formularef{formula:before_tseitin}.
\end{example}

Let $F'$ be the conjunction of clauses forming the problem.
The Davis--Putnam--Logemann--Loveland algorithm (DPLL) decides a propositional formula in CNF (conjunction of clauses) by maintaining a partial assignment of the variables (that is, an assignment to only some of the variables) and \emph{Boolean constraint propagation}: if we have assigned $a=\false,b=\true$ and we have a clause $a \lor \neg b \lor c$, then we can derive $c=\true$. If an assignment satisfies all clauses, then the algorithm terminates with one solution. If it falsifies at least one clause, then there is no solution for our starting partial assignment (thus no solution at all if our starting partial assignment was empty).
If propagation is insufficient to conclude, then the algorithm chooses a variable $x$ and a true value $b$ and extends the assignment with $x=b$; if no solution is found for that assignment, then it \emph{backtracks} and replaces it by $x=\lneg b$. The solver thus constructs a \emph{search tree}.

The practical performance of the solver depends highly on the heuristics for choosing $x$ and~$b$.
Much effort has been put into researching these heuristics, such as \emph{Variable State Independent Decaying Sum} (VSIDS) \cite{Moskewicz:2001:CEE:378239.379017}; understanding why they work well is an active research topic.
The Boolean constraint propagation phase must be implemented very efficiently, using data structures that minimize the traversal of irrelevant data (clauses that will not result in further propagation); e.g. the \emph{two watched literals per clause} scheme \cite[\S 4.5.1.2]{MSLM09HBSAT}.

From a run of the DPLL algorithm concluding to unsatisfiability one can extract a \emph{resolution proof} of unsatisfiability. The proof has the form of a tree whose leaves are some of the original clauses of the problem (constituting an unsatisfiable core) and
whose inner nodes correspond to the choices made during the search.
Each inner node is the application of the \emph{resolution rule}:
knowing $C_1 \lor a$ and $C_2 \lor \lneg{a}$, where $C_1$ and $C_2$ are clauses and $a$ is a choice variable, one can derive $C_1 \lor C_2$, written:
\begin{equation}
\infer{C_1 \lor C_2}{C_1 \lor a & C_2 \lor \lneg{a}} \,.
\end{equation}

\begin{example}
Consider the system of clauses~\ref{system:after_tseitin}. Boolean clause propagation from unit clause $\phi$ simplifies $\lneg{\phi} \lor g$ and $\lneg{\phi} \lor h$ into $g$ and $h$ respectively, and removes clause $\lneg{g} \lor \lneg{h} \lor \phi$.
Since $g$ and $h$ are now $\true$, we can remove clauses $\lneg{e} \lor g$ and $\lneg{f} \lor g$, $b \lor h$, and $c \lor h$, and simplify 
$\lneg{g} \lor e \lor f$ into $e \lor f$ and
$\lneg{h} \lor \lneg{b} \lor \lneg{c}$ into $\lneg{b} \lor \lneg{c}$:

\begin{equation}\label{system:after_tseitin_1bcp}
\begin{array}{c@{\hskip 2em}c@{\hskip 2em}c@{\hskip 2em}c@{\hskip 2em}c}
\lneg{e} \lor a & \lneg{e} \lor \lneg{b} & \lneg{e} \lor \lneg{c}
  & \lneg{a} \lor b \lor c \lor e &
    \lneg{f} \lor b \\
\lneg{f} \lor c &
  \lneg{f} \lor d & \lneg{b} \lor \lneg{c} \lor d \lor f &
  e \lor f & \lneg{b} \lor \lneg{c} \,.\\
\end{array}
\end{equation}

The system no longer has unit clauses to propagate and thus must pick a literal, for instance~$b$.
\ifthenelse{\boolean{fullversion}}{%
Clause $\lneg{e} \lor \lneg{b}$ is simplified into $\lneg{e}$, thus $\lneg{e} \lor \lneg{c}$ and $\lneg{e} \lor a$ are removed and $e \lor f$ is simplified into~$f$.
$\lneg{a} \lor b \lor c \lor e$ and $\lneg{f} \lor b$ are removed.
Clause $\lneg{b} \lor \lneg{c}$ is simplified into $\lneg{c}$.
$\lneg{f} \lor c$ is simplified into $\lneg{f}$. Contradiction.
}{By propagation, the system now reaches a contradiction.}
Since contradiction was reached from assumption $b$, the converse~$\lneg{b}$ must be assumed.
In fact, it is possible to derive the learned clause $\lneg{b}$ by resolution from the set of clauses:

\begin{equation}
\infer{\lneg{b}}{
  \infer{\lneg{b} \lor c}{
    \infer{e \lor c}{e \lor f & \lneg{f} \lor c} &
    \lneg{e} \lor \lneg{b}
  } &
  \lneg{b} \lor \lneg{c}}.
\end{equation}
\end{example}

From any ``unsatisfiable'' run of a DPLL (even in the CDCL variant, see below) solver, a resolution proof can be extracted. This is a fundamental limitation of that approach,%
\ifthenelse{\boolean{fullversion}}{%
since it is known that for certain families of formulas, any resolution proof has exponential size in the size of the formula --- thus any DPLL/CDCL solver will take exponential time.
Any example of such family is the so-called \emph{pigeonhole principle} \cite{DBLP:journals/tcs/Haken85}: a number $m$ of pigeons must be installed in $n$ holes, such that any pigeon is in a hole and no hole contains two pigeon; the variable $a_{ij}$ expresses that pigeon number $i$ is in hole number $j$, and a quadratic number of clauses constrain the problem. For $m > n$ this problem has no solution, by a simple cardinality argument; but resolution proofs for $m = n+1$ will have size exponential in~$n$.%
\footnote{%
This indicates that it is a bad idea to encode cardinality constraints into SAT to be solved by DPLL/CDCL; arithmetic constraints are better left to a specialized arithmetic theory solver.}}{
since it is known that for certain families of formulas, such as the \emph{pigeonhole principle} \cite{DBLP:journals/tcs/Haken85}, any resolution proof has exponential size in the size of the formula --- thus any DPLL/CDCL solver will take exponential time.}

Performance was considerably increased by extending DPLL with \emph{clause learning}, yielding constraint-driven clause learning (CDCL) algorithms \cite{MSLM09HBSAT}. In CDCL, when a partial assignments leads by propagation to the falsification of a clause, the deductions made during this propagation are analyzed to obtain a subset of the partial assignment sufficient to entail the falsification of this clause. This subset yields a conjunction $\hat{x}_1 \land \dots \land \hat{x}_n$ (where $\hat{x}_i$ is either $x_i$ or $\neg x_i$), such that its conjunction with $F'$ is unsatisfiable. In other words, it yields a clause $\neg \hat{x}_1 \lor \dots \lor \neg \hat{x}_n$ that is a consequence of $F'$ (in fact, that clause can be obtained by resolution from~$F'$). This clause can thus be conjoined to the problem $F'$ without changing its set of solutions; but \emph{learning} that clause may help cut branches in the search tree early.

Again, the learned clause appears as the root of a resolution proof whose leaves are clauses of the original problem. Since the same learned clause may be used several times, the final proof appears as a directed acyclic graph (DAG, i.e., a tree with shared sub-branches)\ifthenelse{\boolean{fullversion}}{, as opposed to a tree}{}.
There exist formulas admitting DAG resolution proofs exponentially shorter than the smallest tree resolution proof \cite{Tseitin1983}.

A resolution proof, or a more compact format, may thus be produced during an ``unsatisfiable'' run.
A highly optimized SAT or SMT solver is likely to contain bugs, so it may be desirable to have an independent, simpler, possibly formally verified checker reprocess such as proof~\cite{DBLP:conf/cade/Keller13,DBLP:conf/cpp/ArmandFGKTW11,DBLP:conf/itp/BohmeW10}.

\subsection{DPLL(T)}
\label{sec:DPLLT}
The most common way to deal with atomic propositions inside satisfiability testing is the so-called DPLL($T$) scheme, combining a CDCL satisfiability solver and a decision procedure for conjunctions of propositions from theory~$T$.
A quantifier-free formula $F$ over $T$, say
\begin{equation}
\label{eqn:lra_formula}
(x \geq 0 \lor 2x+y \geq 1) \land (y \geq 0) \land (x + y \leq -1) \,,
\end{equation}
is converted into a propositional formula $F'$ (here $(a \lor b) \land c \land d$) by replacing each atomic proposition by a propositional variable, using a dictionary
(here, $x \geq 0 \mapsto a,\allowbreak 2x+y \geq 1 \mapsto b,\allowbreak
        y \geq 0 \mapsto c,x+y \leq -1 \mapsto d$) and after conversion to canonical form (so that e.g. $x+y \geq 1$ and $2x+2y-2 \geq 0$ are considered the same, and $x+y < 1$ is considered as $\neg (x+y \geq 1)$).
$F'$ realizes a \emph{propositional abstraction} of $F$: any solution of $F$ induces a solution of $F'$, but not all solutions of $F'$ necessarily induce a solution of $F$.

Consider the solution $a=\true,b=\false,c=\true,d=\true$ of $F'$; it corresponds to
\begin{equation}
x \geq 0 \land \neg (2x+y \geq 1) \land y \geq 0 \land x+y \leq -1 \,.
\end{equation}
The inequalities $x \geq 0 \land y \geq 0 \land x+y \leq -1$ have no common solution; in other words, $\neg (a \land c \land d)$ is universally true.
The \emph{theory clause} $\neg a \lor \neg c \lor \neg d$ can be conjoined to~$F'$.
There remains a solution $a=\false,b=\true,c=\true,d=\true$ of $F'$; but it entails the contradiction $2x+y \geq 1 \land y \geq 0 \land x+y \leq -1$.
The theory clause $\lneg b \lor \lneg c \lor \lneg d$ is then conjoined to $F'$. Then the propositional problem becomes unsatisfiable, establishing that $F$ has no solution.
We have therefore refined the propositional abstraction according to spurious counterexamples.

In current implementations, the propositional solver does not wait until a total satisfying assignment is computed to call the decision procedure for conjunctions of theory formulas. Partial assignments, commonly at each decision point in the DPLL/CDCL algorithm, are tested for satisfiability.
In addition, the theory solver may, opportunistically, perform \emph{theory propagation}: if it notices that some asserted constraints imply the truth or falsehood of another known predicate, it can signal it to the SAT solver.
The theory solver should be \emph{incremental}, that is, suited for fast addition or retraction of theory constraints, keeping enough internal state to avoid needless recomputation. The SAT solver should be incremental as well, allowing the dynamic addition of clauses.

Multiple theories may be combined, most often by a variant of the Nelson--Oppen approach~\cite[Ch.~10]{Kroening_Strichman_08}.

\subsection{Linear Real Arithmetic}\label{sec:LRA}
In the case of linear rational, or equivalently real, arithmetic (LRA), the theory solver is typically implemented using a variant \cite{Dutertre_de_Moura_simplex_DPLLT_2006,DBLP:conf/cav/DutertreM06} of the simplex algorithm~\cite{Dantzig_Thapa_97,Schrijver98}.
The atomic (in)equalities from the formula, put in canonical form, are collected; new variables are introduced for the linear combinations of variables that are not of the form $\pm x$ where $x$ is a variable.
For instance, \formularef{eqn:lra_formula} is rewritten as $(x \geq 0 \lor \alpha \geq 1) \land (y \geq 0) \land (\beta \leq -1)$, together with the system of linear equalities $\alpha=2x+y$ and $\beta = x+y$.

The simplex algorithm both maintains a tableau and, for each variable, a current valuation and optional lower and upper bounds. At all times, the simplex tableau contains a system of linear equalities equivalent to this system, such that the variables are partitioned into those (\emph{basic variables}) occurring (each alone) on the left side and those occurring on the right side.
The non-basic variables are assigned one of their bounds, or at least a value between these bounds.
The simplex algorithm tries to fit each basic variable within its bounds; if one does not fit, it makes it non-basic and assigns to it the bound that was exceeded, and selects a formerly non-basic variable to make it basic, through a \emph{pivoting} operation maintaining the equivalence of the system of equalities.

The algorithm stops when either a candidate solution fitting all bounds is found, either one equation in the simplex tableau can be shown to have no solution using interval arithmetic from the bounds of the variables (the interval obtained from the right hand side does not intersect that of the basic variable on the left hand side).
A pivot selection ordering is used to ensure that the algorithm always terminates.
Theory propagation may be performed by noticing that the current tableau implies that some literals are satisfied.

\begin{example}
Consider the system
\begin{equation}
\left\{
\begin{array}{rll}
2 & \leq 2x+y \\
-6 & \leq 2x-3y \\
-1000 & \leq 2x+3y & \leq 18\\
-2 & \leq -2x+5y \\
20 & \leq x+y \,. \\
\end{array}\right.
\end{equation}

This system is turned into a system of equations (``tableau'') and a system of inequalities on the variables:
\begin{equation}\left\{
\begin{array}{lrr@{\qquad}rll}
a = & 2x  & +y  &    2 & \leq a \\
b = & 2x  & -3y &   -6 & \leq b \\
c = & 2x  & 3y  &-1000 & \leq c & \leq 18 \\
d = & -2x & +5y &   -2 & \leq d \\
e = & x   & +y  &   20 & \leq e \,.
\end{array}\right.
\end{equation}

The variables on the left of the equal signs are deemed ``nonbasic'' and those on the right are ``basic''.
The simplex algorithm performs pivoting steps on the tableau, akin to those of Gaussian eliminations, until a tableau such as this one is reached:
\begin{equation}\left\{
\begin{array}{lrr}
e = & 7/16c & -1/16d \\
a = & 3/4c  & -1/4d\\
b = & 1/4c  & -3/4d \\
x = & 5/16c & -3/16d\\
y = & 1/8c  & +1/8d \,.
\end{array}\right.
\end{equation}

Now consider the first equation ($e =$).
By interval analysis, knowing $c \leq 18$ and $d \geq -2$, $-7/16c -1/16d \leq 8$.
Yet $e \geq 20$, thus the system has no solution.
These coefficients $7/16$ and $1/16$ can be applied to the original inequalities constraining $c$ and $d$, with coefficient $1$ for that defining $e$, and the resulting inequalities are summed into a trivially false one:
\begin{equation}
\begin{array}{llrcr}
7/16 & (-2x & -3y) & \geq & -7/16\times 18\\
1/16 & (-2x & +5y) & \geq & -1/16 \times 2\\
1 & x & + y & \geq &  20\\
\hline
& 0 & 0 & \geq & 28 \,.
\end{array}
\end{equation}

\end{example}

By reading nonzero coefficients off the conflicting line of the simplex tableau, one gets a minimal set of contradictory constraints: $d+1$ constraints, corresponding to the nonbasic variable and the basic variables with nonzero multipliers, where $d$ is the dimension of the space.
These multipliers may be presented as an \emph{unsatisfiability witness} to an independent proof checker.

Most SMT solvers implement the simplex algorithm using rational arithmetic.
In most cases arising from verification problems, rational arithmetic can be performed using machine integers, without need for going into extended precision arithmetic~\cite{Leo_rational_simplex_2008}.
A common implementation trick is to use a datatype containing a machine-integer $(\mathit{numerator},\mathit{denominator})$ pair or a pointer to an extended precision rational.%
\footnote{e.g. ZArith \url{https://forge.ocamlcore.org/projects/zarith}}
This approach is however very inefficient in the rare cases where the solver goes a lot into extended precision: the size of numerators and denominators grows fast.

This is why it was proposed to perform linear programming in floating-point arithmetic~\cite{DBLP:conf/sat/FaureNOR08,Monniaux_CAV09,Caminha_Monniaux_PAAR2012,DBLP:conf/fmcad/0001BT14}.%
\footnote{%
The performance with linear programming solvers meant for large industrial instances was however disappointing~\cite{DBLP:conf/sat/FaureNOR08}, due to overhead. Closer integration is needed.}
Because the results of floating-point computations cannot be immediately trusted, some checking is needed. One idea is not to recover floating-point numeric information, but the final partition between basic and nonbasic variables~\cite{Monniaux_CAV09,Caminha_Monniaux_PAAR2012,DBLP:conf/fmcad/0001BT14};
once this partition is known, the tableau is uniquely defined and can be computed by plain linear arithmetic --- Gaussian elimination, or better algorithms, including multimodular \cite[ch.~7]{Stein_modular_forms} or $p$-adic approaches.%
\footnote{As implemented in e.g. Linbox (\url{http://www.linalg.org/}), IML (\url{https://cs.uwaterloo.ca/~astorjoh/iml.html}) \cite{Chen:2005:BBC:1073884.1073899} and SageMath (\url{http://www.sagemath.org/}).}
It is then easy to check the alleged conflicting line, in exact precision.

In some cases, linear arithmetic reasoning may be used to prove the unsatisfiability of polynomial problems. One approach is to expand polynomials and consider all monomials as independent variables (e.g. $xy^2$ is replaced by a fresh unknown $v_{xy^2}$). A refinement \cite{DBLP:conf/vmcai/MarechalFKMP16} is to consider lemmas stating that if two polynomials are nonnegative, then so is their product: e.g. $x - 1 \geq 0 \land y - 2 \geq 0 \implies v_{xy} - 2x - y + 2 \geq 0$.%
\footnote{%
One can in fact prove a form of completeness of that approach when the problem contains linear constraints defining a bounded polyhedron, and one nonlinear constraint: if such a problem is unsatisfiable, then this can be proved by going to a sufficiently high degree of products. This follows from Krivine--Handelman's theorem \cite{Krivine_64,handelman_representing_1988}.}
Because the set of such products has size exponential in the maximal degree, heuristics are used to pick the most promising ones.
Experiments have shown this approach to be competitive, even with a rudimentary and sub-optimal connection between linear SMT-solver and nonlinear reasoning.

Some earlier solvers (e.g. CVC3) solver linear real arithmetic by Fourier-Motzkin elimination~\cite{Fourier_1924}. This approach is generally not considered efficient, since Fourier-Motzkin elimination tends to generate many redundant constraints, which then may need to be eliminated by linear programming, which defeats the purpose of avoiding using the simplex algorithm.

\subsection{Linear Integer Arithmetic}
\label{sec:LIA}
In the case of linear integer arithmetic, the scheme generally used is the same as the one generally used for integer linear programming:
the solver first attempts solving the rational relaxation of the problem (nonstrict inequalities are kept, strict inequalities $x < e$ are rewritten as $x \leq e-1$). If there is no solution over the rationals, there is no integer solution.
If a rational solution is found, and has only integral coefficients (say, $(x,y,z)=(0,1,2)$), then the problem is decided.

If the proposed solution has non-integral coefficients (say, $(x,y,z)=(\frac{1}{3},0,1)$), then it is excluded by a constraint removing not only that spurious solution but a whole chunk of them.
Traditional approaches include
\begin{inparaenum}[i)]
\item \emph{branch-and-bound} \cite[Sec.~24.1]{Schrijver98}: add a lemma excluding one segment of non-integral values of the fractional unknowns (here, $x \leq 0 \lor x \geq 1$);  branching is however not guaranteed to terminate in general~\cite{DBLP:conf/fmcad/0001BT14}.
\item \emph{Gomory cuts} \cite[Ch.~23]{Schrijver98}
\item \emph{branch-and-cut} \cite{Mitchell_branch_and_cut}, a combination of both of the above
\item \emph{cuts from proofs} or \emph{extended branches} \cite{DBLP:journals/fmsd/DilligDA11}, which can generate e.g. $x \leq z \lor x \geq z+1$.
\end{inparaenum}

The full integer linear decision procedure can be encapsulated and only export theory lemmas and theory propagation, just as the rational linear procedure, or export the branching lemma to the SMT solver, as a learned clause, so as to allow propositional reasoning over it.

An alternative to linear programming plus branching and/or cuts is Pugh's Omega test \cite{Pugh:1991:OTF:125826.125848}, which may also be used to simplify constraints. This test is based on Fourier-Motzkin elimination \cite{Fourier_1924}, with the twist that, due to divisibility constraints, it may need to enumerate cases up to the least common multiple of the divisors.

\subsection{Exponential Behavior Due to Limited Predicate Vocabulary}
\label{sec:exponential_WCET}
\begin{example}\label{ex:exponential_WCET}
Let $n > 0$ be a constant integer. Let $(t_i)_{0 \leq i \leq n}$, $(x_i)_{0 \leq i < n}$ and $(y_i)_{0 \leq i < n}$ be real unknowns (or rational or integer). Let
\begin{align}
D_i \defn& (x_i - t_i \leq 2) \land (y_i - t_i \leq 3) \land
   \left( (t_{i+1} - x_i \leq 3) \lor (t_{i+1} - y_i \leq 2) \right) \,,\\
P_n \defn& \bigwedge_{i=0}^{n-1} D_i \land t_n - t_0 > 5n \,.
\end{align}

These formulas are known as ``diamond formulas'' since they correspond to paths in a difference graph composed of ``diamonds'':
\begin{center}
\begin{tikzpicture}[node distance=4em]
\node (t0) {$t_0$};
\node[above right of=t0] (x0) {$x_0$};
\node[below right of=t0] (y0) {$y_0$};
\node[below right of=x0] (t1) {$t_1$};
\path[->] (t0) edge node[above left]{$2$} (x0);
\path[->] (t0) edge node[below left]{$3$} (y0);
\path[->] (x0) edge node[above right]{$3$} (t1);
\path[->] (y0) edge node[below right]{$2$} (t1);

\node[above right of=t1] (x1) {$x_1$};
\node[below right of=t1] (y1) {$y_1$};
\node[below right of=x1] (t2) {$t_2$};
\path[->] (t1) edge node[above left]{$2$} (x1);
\path[->] (t1) edge node[below left]{$3$} (y1);
\path[->] (x1) edge node[above right]{$3$} (t2);
\path[->] (y1) edge node[below right]{$2$} (t2);

\node[right of=t2, node distance=8em] (tnm1) {$t_{n-1}$};
\path[->,dotted] (t2) edge (tnm1);

\node[above right of=tnm1] (xnm1) {$x_{n-1}$};
\node[below right of=tnm1] (ynm1) {$y_{n-1}$};
\node[below right of=xnm1] (tn) {$t_n$};
\path[->] (tnm1) edge node[above left]{$2$} (xnm1);
\path[->] (tnm1) edge node[below left]{$3$} (ynm1);
\path[->] (xnm1) edge node[above right]{$3$} (tn);
\path[->] (ynm1) edge node[below right]{$2$} (tn);
\end{tikzpicture}
\end{center}

To a human, it is obvious that $D_i \Rightarrow t_{i+1} \leq t_i + 5$ and thus $P_n$ is unsatisfiable.
A DPLL(T) solver, however, proceeds by elimination of contradictory conjunctions of atoms from the original formula.
Any contradictory conjunction of atoms from $P_n$ must include a conjunction of the form
$\bigwedge_{i=0}^{n-1} F_i \land  t_n - t_0 > 5n$ where $F_i$ is either
$(x_i - t_i \leq 2) \land (t_{i+1} - x_i \leq 3)$ or
$(y_i - t_i \leq 3) \land (t_{i+1} - y_i \leq 2)$.
There are an exponential number of such conjunctions, and a DPLL(T) solver has to block them by theory lemmas one by one.
\end{example}

In other words, the proof system used by a DPLL(T) solver is sufficient to prove that a ``diamond formula'' is unsolvable, but needs exponential proofs for doing so.
Any pure DPLL(T) solver, whatever its heuristics and implementation, must thereof run in exponential time on this family of formulas.
This motivated the study of algorithms capable of inferring lemmas involving new atoms (Sec.~\ref{sec:MCSAT}).

\ifthenelse{\boolean{fullversion}}{%
Diamond formulas are not a mere academic curiosity.
In \emph{worst-case execution time} (WCET) analysis~\cite{Henry_et_al_WCET_LCTES2014}, a variable ($t_i$, $x_i$, $y_i$) within a node represents the arrival time of the program (in clock cycles) at that location, and one seasrches for the maximum of $t_n$ --- in our example formula $P_n$ establishes that the worst-case execution time (WCET) of the program is at most~$5n$.
In this case, the DPLL(T) solver in effect needs to explore individually all $2^n$ paths in the program to reach that conclusion.
This makes binary search (Sec.~\ref{sec:optimization}) for the optimum intractable, with queries becoming exponentially more expensive as the solver narrows the interval around the bound.
Worse, in real applications, the formula however is much more complicated than $P_n$, since it also encodes the semantics of the programming constructs (Sec.~\ref{sec:program_analysis}), which may exclude certain paths (e.g. some block of code may never be executed if some other block is executed).

}{Diamond formulas are simplifications of formulas occurring in e.g. worst-case execution time and scheduling applications.}
The solution proposed in \cite{Henry_et_al_WCET_LCTES2014} was to pre-compute upper bounds $t_j - t_i \leq B_{ij}$ on the difference of arrival times between $i$ and $j$ (or, equivalently, the total time spent in the program between $i$ and $j$) and conjoin these bounds to the problems.
These bounds are logically implied by the original problem, and thus the set of solutions (valid execution traces with timings) does not change; but the resulting formula is considerably more tractable.
The lemmas $t_j - t_i \leq B_{ij}$ and $t_k - t_j \leq B_{ik}$ allow the solver to avoid exploring many combinations of paths $i \rightarrow j$ and $j \rightarrow k$:
for instance, if one searches for a path such that $t_k - t_i \geq 100$, it is known that $t_k - t_j \leq 40$, and the solver explores a path $i \rightarrow j$ such that $t_j - t_i \leq 42$ on this path, then the solver can immediately cut the search without exploring the paths $j \rightarrow k$ in detail.

\subsection{Uninterpreted Functions and Arrays}
\label{sec:UF}
There exists several variants of how to decide uninterpreted functions (UF) in combination with other theories~\cite[Ch.~4]{Kroening_Strichman_08}; we shall expose only one approach here.
A quantifier-free formula (e.g. $f(x)\neq f(y) \land x=z+1 \land z=y-1$) is rewritten so that each application of an uninterpreted function is replaced by a fresh variable (e.g. $f_x \neq f_y \land x=z+1 \land z=y-1$), several identical applications getting the same variable.
A solution in $x,y,z,f_x,f_y$ is sought. If $x=y$ but not $f_x \neq f_y$ in that solution, the implication $x=y \Rightarrow f_x=f_y$ is conjoined to the problem.
Again, this is a counterexample-guided refinement of the theory.

\begin{example}
$f(x)\neq f(y) \land x=z+1 \land z=y-1$, where $x,y,z \in \ZZ$ and $f: \ZZ \rightarrow \ZZ$, has no solution because $x=z+1 \land z=y-1$ implies that $x=y$, and it is then impossible that $f(x) \neq f(y)$.
One may establish this by solving $f_x \neq f_y \land x=z+1 \land z=y-1$, getting $(x,y,z,f_x,f_y) = (1,1,0,0,1)$, noticing the conflict between $x=y$ and $f_x \neq f_y$ and conjoining $x=y \Rightarrow f_x=f_y$.
\end{example}

\emph{Arrays} are ``functionally updatable'' uninterpreted functions \cite[Ch.~7]{Kroening_Strichman_08}:\\
$\mathit{update}(f,x_0,y_0)$ is the function mapping $x \neq x_0$ to $f[x]$ and $x_0$ to $y_0$.

\section{Natural-Domain SMT}
\label{sec:natural_domain}
In DPLL(T) there is a fundamental difference between propositional and other kinds of unknowns: the second are never dealt with directly during the search process. In contrast, in \emph{natural-domain} SMT, one directly constrains and assigns to numeric unknowns during the search. After initial attempts \cite{McMillan_et_al_CAV09,DBLP:conf/formats/Cotton10},
two main directions arose.

\subsection{Abstract CDCL (ACDCL)}
The DPLL approach is to assign to each unknown (propositional variable) one of $\true$, $\false$, and ``undecided'' --- that is, a non-empty subset of the set of possible values $\{ \true, \false \}$.
Initially, all variables are assigned to ``undecided''. Then, the Boolean constraint propagation phase uses each individual clause as a constraint over its literals: if all literals except for one are assigned to~$\false$, then the last one gets assigned to~$\true$. In other words, information known about some variables leads to information on other variables linked by the same constraint.
If the information derived is that some variable cannot be assigned some value (``contradiction''), then it means the problem is unsatisfiable.
In most cases, however, a contradiction cannot be derived by only the initial pass of propagation.
In that case, the system picks an undecided variable and splits the search between the $\true$ and $\false$ cases. Several splits may be needed, thus the formation of a search tree. If a contradiction is derived in a branch, that branch is closed and the system backtracks to an earlier level.

That approach may be extended to variables lying within an arbitrary domain $D$, say, the real numbers or the floating-point numbers. The system maintains for each variable an assignment to a subset of $D$ (several types of variables may be used simultaneously, there may therefore be several~$D$), chosen among an \emph{abstract domain}%
\footnote{Following the terminology of \emph{abstract interpretation}; see \cite{DBLP:journals/fmsd/BrainDGHK14} for more.}
$\abstr{D}$ of subsets of $D$; say, for numeric variables, $\abstr{D}$ may be the set of closed intervals of $D$.
Constraints may now constrain variables of different types, and each constraint acts as a propagator of information.
For instance, if there is a constraint $x = y+z$, and $x$ is currently assigned the interval $[1,+\infty)$ and $y$ the interval $[4,10]$, then, applying $z = x-y$, one can derive $x \in [-9,+\infty)$: the current interval for $x$ may thus be refined.

Note that, for soundness, it is not important that the information propagated should be optimally precise, as long as it contains the possible values: in the above example, it would be sound to propagate $x \in [-9.1,+\infty)$ --- but unsound to derive $xx \in [-8.99,+\infty)$.
In the case of interval propagation for $D = \RR$, one sound way to implement it is using floating-point interval arithmetic with directed rounding: the upper bound of an interval is rounded towards $+\infty$, the lower bound towards $-\infty$.

ACDCL also applies clause learning, but in a more general manner than CDCL \cite[Sec.~5]{DBLP:journals/fmsd/BrainDGHK14}.
Consider $F \defn y = x \land z=x\cdot y \land z\leq -1$
and a search context with $x \leq -4$. Then, by interval propagation, $y \leq -4$, and $z \geq 16$, which contradicts $z \leq -1$. CDCL-style clause learning would learn that $x \leq -4$ contradicts $F$, and thus learn the clause $\neg (x \leq -4) \equiv x > -4$.
But there is a weaker reason why such choice of $x$ contradicts $F$: $x < 0$ is sufficient to ensure contradiction; the solver can exclude a larger part of the search space by learning the clause $\neg (x < 0) \equiv x \geq 0$.
Generalizing the reasons for a contradiction is a form of \emph{abduction}.
One difficulty is that there may be no weakest generalization expressible in the abstract domain: for instance, the choices $x \geq 10$ and $y \geq 10$ contradict the constraint $x+y < 10$, but $x \geq 0 \land y \geq 10$, $x \geq 5 \land y \geq 5$ and $x \geq 10 \land y \geq 0$ are three incomparable generalizations of the contradiction (leading to three clauses $x < 0 \lor y < 10$ etc.), which are optimal in the sense that if one fixes the interval for $x$ (resp.~$y$), the interval for $y$ (resp.~$x$) is the largest that still ensures contradiction.

\subsection{Model-Constructing Satisfiability Calculus (MCSAT)}
\label{sec:MCSAT}
In DPLL(T) \begin{inparaenum}[i)]
\item only propositional atoms (including Boolean unknowns) are assigned during the search
\item the set of atoms considered does not change throughout the search (this may cause exponential behavior, see Sec.~\ref{sec:exponential_WCET}
\item when the search process, after assigning $b_1,\dots,b_n$ concludes that it is impossible to assign a Boolean value to an atom $b_{n+1}$, it derives a learned clause over a subset of $b_1,\dots,b_n$ that excludes the current assignment but also, hopefully, many more.
\end{inparaenum}
In contrast, in \emph{model-constructing satisfiability calculus} (MCSAT) \cite{DBLP:conf/vmcai/MouraJ13}, both propositional atoms and numeric unknowns get assigned during the search, and new arithmetic predicates are generated through learning.

\subsubsection{Linear Real Arithmetic}
Assume variables $x_1,\dots,x_n$ have been assigned values $v(x_1),\dots,v(x_n)$ in the current branch of the search, and that two atoms
$x_{n+1} \leq a$ and
$x_{n+1} \geq b$, where $a$ and $b$ are linear combinations of variables other than $x_{n+1}$, have been assigned to $\true$, such that $b > a$ in the assignment $v$;
then it is impossible to pick a value for $x_{n+1}$ in that assignment.
In fact, it is impossible to pick a value for it in \emph{any} assignment such that $b > a$.

Assignments that conflict for the same reason are eliminated by a \emph{Fourier-Motzkin elimination} \cite{Fourier_1924} elementary step, valid for all $x_1,\dots,x_{n+1}$:
\begin{equation}
\neg x_{n+1} \leq a \lor \neg x_{n+1} \geq b \lor a \geq b \,.
\end{equation}

\begin{example}\label{ex:MCSAT_WCET}
Consider Ex.~\ref{ex:exponential_WCET} with $n=3$.
The solver has clauses
$x_i - t_i \leq 2$, $y_i - t_i \leq 3$,
$t_{i+1} - x_i \leq 3 \lor t_{i+1} - y_i \leq 2$ for $0 \leq i <3$,
and $t_0 = 0$, $t_3 \geq 16$.

The solver picks
$t_0 \mapsto 0$, $t_1-x_0 \leq 3 \mapsto \true$, $x_0 \mapsto 0$,
$t_1 \mapsto 0$, $t_2-x_1 \leq 3 \mapsto \true$, $x_1 \mapsto 0$,
$t_2 \mapsto 0$, $t_3-x_2 \leq 3 \mapsto \true$, $x_2 \mapsto 0$.
But then, there is no way to assign $t_3$, because of the current assignment $x_2 \mapsto 0$ and the inequalities $t_3-x_2 \leq 3$ and $t_3 \geq 16$.
The solver then learns by Fourier-Motzkin:
\begin{equation}
\neg(t_3 \geq 16) \lor \neg(t_3-x_2 \leq 3) \lor x_2 \geq 13 \,.
\end{equation}
which may in fact be immediately simplified by resolution with the original clause $t_3 \geq 16$ to yield $\neg(t_3-x_2 \leq 3) \lor x_2 \geq 13$.
The assignment to $x_2$ is retracted.

But then, there is no way to assign $x_2$, because of the current assignment $t_2 \mapsto 0$ and the inequality $x_2 - t_2 \leq 2$.
The solver then learns by Fourier-Motzkin:
\begin{equation}
\neg(x_2 \geq 13) \lor \neg(x_2-t_2 \leq 2) \lor t_2 \geq 11 \,.
\end{equation}
By resolution, $\neg(t_3-x_2 \leq 3) \lor t_2 \geq 11$.
The truth assignment to $t_3-x_2 \leq 3$ is retracted.

At this point, the solver has
$t_0 \mapsto 0$, $t_1-x_0 \leq 3 \mapsto \true$, $x_0 \mapsto 0$,
$t_1 \mapsto 0$, $t_2-x_1 \leq 3 \mapsto \true$, $x_1 \mapsto 0$,
$t_2 \mapsto 0$, $t_3-x_2 \leq 3 \mapsto \false$.
By similar reasoning in that branch,
the solver derives $t_3-x_2 \leq 3 \lor t_2 \geq 11$.
By resolution between the outcomes of both branches, one gets $t_2 \geq 11$.

By similar reasoning, one gets $t_1 \geq 6$ and then $t_0 \geq 1$, but then there is no satisfying assignment to $t_0$. The problem has no solution.

In contrast to the exponential behavior of DPLL(T) on Ex.~\ref{ex:exponential_WCET}, MCSAT has linear behavior: each branch of each individual disjunction is explored only once, and the whole disjunction is then summarized by an extra atom.
\end{example}

The dynamic generation of new atoms by MCSAT, as opposed to DPLL(T), creates two issues.
\begin{inparaenum}[i)]
\item If infinitely many new atoms may be generated, termination is no longer ensured. One can  ensure termination by restricting the generation of new atoms to a \emph{finite basis} (this basis of course depends on the original formula);
this is the case for instance if the numeric variables $x_1,\dots,x_n$ are always assigned in the same order, thus the generated new atoms are results of Fourier-Motzkin elimination of $x_n$, then of $x_{n-1}$ etc. down to $x_2$.%
\footnote{Successive applications of Fourier-Motzkin may lead to very large sets of predicates, thus this argument seems of mostly theoretical interest.}
In practice, the interest of being able to choose variable ordering trumps the desire to prove termination.
\item Since many new atoms and clauses are generated, some garbage collection must be applied, as with learned clauses in a CDCL solver.
\end{inparaenum}

Implementation-wise, note that, like a clause in CDCL, a linear inequality is processed only when all variables except for one are assigned. Similar to \emph{two watched literals per clause}, one can apply \emph{two watched variables per inequality}.

\subsubsection{Nonlinear Arithmetic (NRA)}
The MCSAT approach can also be applied to polynomial real arithmetic.
Again, the problem is: assuming a set of polynomial constraints over $x_1,\dots,x_n,x_{n+1}$ have no solution over $x_{n+1}$ for a given valuation $v(x_1),\dots,v(x_n)$, how can we explain this impossibility by a system of constraints over $x_1,\dots,x_n$ that excludes $v(x_1),\dots,v(x_n)$ and hopefully many more?

Jovanovi\'c and de Moura \cite{Jovanovic_de_Moura_IJCAR2012} proposed applying a modified version of Collin's \cite{Collins_CAD} projection operator in order to perform a partial \emph{cylindrical algebraic decomposition}.
In that approach, known as NLSAT, one additional difficulty is that assignments to variables may refer to algebraic reals, and thus the system needs to compute to compute over algebraic reals, including as coefficients to polynomials.
It is yet unknown whether this approach could benefit from using other projection operators such as Hong's \cite{Hong:1990:IPO:96877.96943} or McCallum's~\cite{McCallum98}.

\section{Beyond Quantifier-Free Decidability}
\label{sec:beyond}
\subsection{Quantifiers}
\label{sec:quantifiers}
\subsubsection{Quantifier Elimination by Virtual Substitution}
In the case of some theories, such as linear real arithmetic, a finite sequence of instantiations can be produced such that $F \defn \forall x~ P(x)$ is equivalent to $\bigwedge_{i=1}^n P(v_i)$; note that the $v_i$ are not constants, but functions of the free variables of~$F$, obtained by analyzing the atoms of~$P$.
Because this approach amounts to substituting expressions into the quantified variable, it is called \emph{substitution}, or \emph{virtual substitution} if appropriate data structures and algorithms avoid explicit substitution.
Examples of substitution-based methods include Cooper's \cite{Cooper72} for linear integer arithmetic, Ferrante \& Rackoff's \cite{FerranteRackoff75} and Loos \& Weisfpenning's \cite{LoosWeispfenning93} methods for linear real arithmetic.

\begin{example}
Consider $\forall y~ (y \geq x \Rightarrow y \geq 1)$. Loos \& Weisfpenning's method collects the expression to which $y$ is compared (here, $x$ and $1$) and then substitutes them into $y$. For each expression $e$, one must also substitute $e + \epsilon$ where $\epsilon$ is infinitesimal,%
\footnote{$x \geq K + \epsilon$ with $K$ real means $x > K$.}
and also substitute $-\infty$ (equivalently, one can substitute $e-\epsilon$ for each expression, and also $+\infty$). The result is therefore
\begin{equation}
\bigwedge_{e \in \{x, x+\epsilon, 1, 1+\epsilon, -\infty\}} e \geq x \Rightarrow e \geq 1
\end{equation}
or, after expansion and simplification, $x \geq 1$.
\end{example}

We thus have \emph{eliminated} the quantifier; by recursion over the structure of a formula and starting at the leaves, we can transform any formula of linear real arithmetic into an equivalent quantifier-free formula.%
\footnote{%
In the case of linear integer arithmetic, we need to enrich the language of the output formula with constraints of divisibility by constants: e.g. $\exists x~ y = 2x$ is equivalent to quantifier-free $2 \mid x$.}

In these eager approaches, the size of the substitution set may grow quickly (especially for linear integer arithmetic, which may involve enumerating all cases up to the least common multiple of the divisibility constants). For this reason, lazy approaches were proposed where the substitutions are generated from counterexamples, in much the same way that learned lemmas are generated in DPLL(T) \cite{Phan_Bjorner_Monniaux_SMT2012,Bjorner_IJCAR10}. 
For a formula $A(\ve{x}) \land \forall y~ B(\ve{x},y)$, the system first solves $A(\ve{x})$ for a solution $\ve{x}_0$, then checks whether there exists $y$ such that $\neg B(\ve{x}_0,y)$; if so, such an $y_0$ is generalized into one of the possible substitutions $S_1(\ve{x})$ and the system restarts by solving $A(\ve{x}) \land B(\ve{x},S_1(\ve{x})$. The process iterates until a solution is found or the substitutions accumulated block all solutions for $\ve{x}$; termination is ensured because the set of possible symbolic substitutions is finite. 
Note that a full quantifier elimination is not necessary to produce a solution.

\subsubsection{Quantifier Elimination by Projection}
In case a quantifier elimination, or \emph{projection}, algorithm, is available for conjunctions of constraints, as happens with linear real arithmetic,%
\footnote{This amounts to projection of convex polyhedra, for which there exist algorithms based on conversion to generators (vertices), Fourier-Motzkin elimination and pruning, or parametric linear programming, among others~\cite{Fouilhe_PhD}.}
one can, given a formula $\exists \ve{y}~F(\ve{x},\ve{y})$, find a conjunction $C_1 \Rightarrow F$, project $C_1$ over $\ve{x}$ as $\pi(C_1)$, conjoin $\neg \pi(C_1)$ to $F$ and repeat the process (generating $C_2$ etc.) until the $F$ becomes unsatisfiable~\cite{Monniaux_LPAR08}. $\bigvee_i C_i$ is then equivalent to $\exists \ve{y}~F$.
Again, this process may be made lazier, for nested quantification in particular \cite{Phan_Bjorner_Monniaux_SMT2012,Monniaux_CAV10}.

\subsubsection{Instantiation Heuristics}
The addition of quantifiers to theories (such as linear integer arithmetic plus uninterpreted functions) may make them undecidable.
This does not however deter designers of SMT solvers from attempting to have them decide as many formulas as possible.
A basic approach is quantifier instantiation by E-matching.
If a formula in negation normal form contains a subformula $\forall x~ P(x)$, then this formula is replaced by a finite instantiation
$\bigwedge_{i=1}^n P(v_i)$. The $v_i$ are extracted from the rest of the formula, possibly guided by counterexamples.
This approach is not guaranteed to converge: an infinite sequence of instantiations may be produced for a given quantifier.
In the case of \emph{local theories}, one can however prove termination. 

\subsection{Craig Interpolation}
\label{sec:interpolants}
The following conjunction is satisfiable if and only if it is possible to go from a model $\ve{x}_0$ of $A$ to a model $\ve{x}_n$ of $B$ by a sequence of transitions $(\tau_i)_{1 \leq i \leq n}$:
\begin{equation}\label{formula:Craig_Unsat}
A(\ve{x}_0) \land \tau_1(\ve{x}_0,\ve{x}_1) \land \dots \land \tau_n(\ve{x}_{n-1},\ve{x}_n) \land B(\ve{x}_n) \,.
\end{equation}
In program analysis, $A$ typically expresses a precondition, $\neg B$ a postcondition, $\ve{x}_i$ the variables of the program after $i$ instruction steps, and $\tau_i$ the semantics of the $i$-th instruction in a sequence, and the formula is unsatisfiable if and only if $B$ is always true after executing that sequence of instruction starting from~$A$.

A hand proof of unsatisfiability would often consist in exhibiting predicates $I_1(\ve{x}_1),\dots,I_{n-1}(\ve{x}_{n-1})$, such that, posing $I_0=A$ and $I_n=B$, for all $0 \leq i < n$,
\begin{equation}\label{formula:local_inductiveness}
\forall \ve{x}_i,\ve{x}_{i+1}~ I_i(\ve{x}_i) \land \tau_{i+1}(\ve{x}_i,\ve{x}_{i+1}) \Rightarrow I_{i+1}(\ve{x}_{i+1}) \,,
\end{equation}
along with proofs of these \emph{local inductiveness} implications.%
\footnote{%
In program analysis, this corresponds to stating ``after the first instruction, the program variables satisfy $I_1$, but then if one executes the second instruction from $I_1$, the program variables then satisfy $I_2$\dots'', and $\{ I_i \} ~ \tau_i~ \{ I_{i+1} \}$ constitute Hoare triples.}

A SMT-solver, in contrast, produces a monolithic proof of unsatisfiability of \formularef{formula:Craig_Unsat}: it mixes variables from different $\ve{x}_i$, that is, in program analysis, from different times of the execution of the program.
It is however possible to obtain instead a sequence $I_i$ satisfying \formularef{formula:local_inductiveness} by post-processing that proof \cite{McMillan_TCS05,Christ_PhD,DBLP:conf/tacas/ChristHN13}.

In any theory admitting quantifier elimination, such a sequence must exist:
\begin{equation}
I_{i+1} \equiv \exists \ve{x}_i~ I_i(\ve{x}_i) \land \tau_{i+1}(\ve{x}_i,\ve{x}_{i+1})
\end{equation}
defines the strongest sequence of valid interpolants; the weakest is:
\begin{equation}
I_i \equiv \forall \ve{x}_{i+1}~ \tau_{i+1}(\ve{x}_i,\ve{x}_{i+1}) \Rightarrow I_{i+1}(\ve{x}_{i+1}) \,.
\end{equation}
The strongest sequence corresponds to computing exactly the sequence of sets of states reachable by $\tau_1$, then $\tau_2 \circ \tau_1$ etc. from $A$.

\emph{Binary interpolation} consists in: given $A$ and $B$, produce $I$ such that
\begin{equation}
\forall \ve{x}_0,\ve{x}_1,\ve{x}_2~ A(\ve{x}_0,\ve{x}_1) \Rightarrow I(\ve{x}_1) \Rightarrow B(\ve{x}_1,\ve{x}_2) \,,
\end{equation}
in which case, if the theory admits quantifier elimination, $\exists \ve{x}_0~ A(\ve{x}_0,\ve{x}_1)$ and $\forall \ve{x}_2~ B(\ve{x}_1, \ve{x}_2)$ are respectively the strongest and weakest interpolant, and any $I$ in between ($\exists \ve{x}_0~ A(\ve{x}_0,\ve{x}_1) \Rightarrow I \Rightarrow \forall \ve{x}_2~ B(\ve{x}_1, \ve{x}_2)$) is also an interpolant.

\ifthenelse{\boolean{fullversion}}{
Note, however, that there exist theories for which there is no quantifier elimination but Craig interpolation is possible. 
Also, the interpolants obtained by quantifier elimination may be unduly complicated, or also ill-suited for certain purposes; let us see why.
Let us take $A(x,y) \defn x=y=0$, $B \defn y \leq 2000$, and
$\tau\left((x,y),(x',y')\right) \defn x \leq 1000 \land x'=x+1 \land y'=y+1$.
We wish to show that any $(x,y)$ reachable from $A$ from any number of application of $\tau$ satisfies~$B$.
One idea is to show that $\neg B$ is unreachable after $1$ step of $\tau$, $2$ steps of $\tau$\dots and to generalize the argument to any number of steps.

Witness that $A(x_0,y_0) \land \tau\left((x_0,y_0),(x_1,y_1)\right) \Rightarrow B(x_1,y_1)$, and that
$A(x_0,y_0) \land
 \tau\left((x_0,y_0),(x_1,y_1)\right) \land
 \tau\left((x_1,y_1),(x_2,y_2)\right) \Rightarrow B(x_2,y_2)$.
There are several possible interpolants $I_1(x_1,y_1)$: the strongest is
$\exists x0,y0~ x_0=0 \land =y_0=0 \land x_0 \leq 1000 \land x_1=x_0+1 \land y_1=y_0+1
\equiv x_1=1 \land y_1=1$,
the weakest is
$\forall x_2,y_2~ x_1 \leq 1000 \land x_2=x_1+1 \land y_2=y_1+1 \Rightarrow y_2 \leq 2000
\equiv y_1 \leq 1999$.
Note that neither of these invariants is inductive with respect to $\tau$: neither satisfy $\forall x,y,x',y'~ I_1(x,y) \land \tau\left((x,y),(x',y')\right) \Rightarrow I_1(x',y')$.
Yet the following interpolant is inductive: $I_1(x,y) \defn x=y$: it satisfies $A \Rightarrow I_1$ and $I_1 \land \tau \Rightarrow B$.}

One of the main uses of Craig interpolation in program analysis is to synthesize inductive invariants, for instance by counterexample-guided abstraction refinement in predicate abstraction (CEGAR) \cite{DBLP:conf/cav/McMillan06}
or property-guided reachability (PDR). 
Interpolants obtained by quantifier elimination are too specific (\emph{overfitting}): for instance, strongest interpolants exactly fit the set of states reachable in $1,2,\dots$ steps.
It has been argued that interpolants likely to be useful as inductive invariants should be ``simple'' --- short formula, with few ``magical constants''. A variety of approaches have been proposed for getting such interpolants \cite{DBLP:conf/cav/SharmaNA12,DBLP:conf/cav/AlbarghouthiM13,DBLP:conf/tacas/UnnoT15} or to simplify existing interpolants \cite{DBLP:conf/popl/HoderKV12}.

\begin{figure}[tb]
\begin{center}
\includegraphics[scale=0.5]{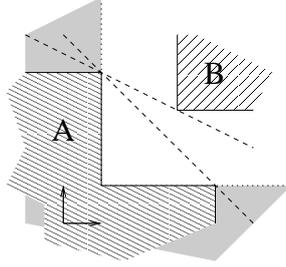}
\end{center}
\vspace{-5mm}

\caption{Binary interpolation in linear arithmetic. The hashed areas represent $A$ and $B$ (Eq.~\ref{eqn:interpolation}) respectively. A possible interpolant $I$ between $A$ and $B$ ($A \Rightarrow I$, $I \Rightarrow \neg B$) is the grey area $x \leq 1 \lor y \leq 1$. The dashed lines define two other possible interpolants, $x+y \leq 5$ and $x+2y \leq 9$.}
\label{fig:binary_interpolation}
\end{figure}

\begin{example}\label{ex:interpolation}
Consider the interpolation problem $A \Rightarrow I$, $I \Rightarrow \neg B$ (Fig.~\ref{fig:binary_interpolation}):
\begin{equation}
\begin{array}{ll@{\qquad}ll}
A_1 & \defn x\leq 1 \land y\leq 4 &
A_2 & \defn x\leq 4 \land y\leq 1\\
A   & \defn A_1 \lor A_2  &
B   & \defn x\geq 3 \land y\geq 3 \,.
\end{array} \label{eqn:interpolation}
\end{equation}

\soft{SMTInterpol}\footnote{\soft{SMTInterpol} 2.1-31-gafd0372-comp}
and \soft{MathSAT}\footnote{\soft{MathSAT} 5.3.10}
produce
$I \defn x \leq 1 \lor y \leq 1$.
This is due to the way these tools produce interpolants from DPLL(T) proofs of unsatisfiability.
On this example, a DPLL(T) solver will essentially analyze both branches of $A_1 \lor A_2$. The first branch yields $A_1 \Rightarrow \neg B$.
Finding $I_1$ such that $A_1 \Rightarrow I_1$ and $I_1 \Rightarrow \neg B$ amounts to finding a separating hyperplane between these two convex polyhedra; $I_1 \defn x \leq 1$ works.
Similarly, $I_2$ such that $A_1 \Rightarrow I_2$ and $I_2 \Rightarrow \neg B$ can be $I_2 \defn y \leq $. $I_1 \lor I_2$ is then produced as interpolant.

Yet, a search for a single separating hyperplane may produce $x+2y \leq 9$, or $x+y \leq 5$. The second hyperplane may seem preferable according to a criterion limiting the magnitude of integer constants.
\end{example}

\ifthenelse{\boolean{fullversion}}{%
As seen on the above example, if a theory admits interpolation between conjunctions of atomic propositions (find $I(\ve{y})$ such that $A(\ve{x},\ve{y}) \Rightarrow I(\ve{y})$ and $I(\ve{y}) \Rightarrow \neg B(\ve{y},\ve{z})$ universally, where $A$ and $B$ are conjunctions of atoms), then it admits interpolation between arbitrary predicates. One simply has to put $A$ and $B$ (such that $A \land B$ is unsatisfiable) into disjunctive normal form $A_1 \lor \dots \lor A_m$ and $B_1 \lor \dots \lor B_n$, and find $I_{i,j}$ such that $A_i(\ve{x},\ve{y}) \Rightarrow I_{i,j}(\ve{y})$ and $I_{i,j}(\ve{y},\ve{z}) \Rightarrow \neg B_j(\ve{z})$ universally.
Then $I = \bigvee_i \bigwedge_j I_{i,j}$ is such an interpolant.

}{It is easy to see that if one can find interpolants for arbitrary conjunctions $A,B$ such that $A \Rightarrow \neg B$, one can find them between arbitrary quantifier-free formulas, by putting them into DNF.}
Because such a procedure would be needlessly costly due to disjunctive normal forms, the usual approach is to post-process a DPLL(T) proof that $A(\ve{x},\ve{y}) \land B(\ve{y},\ve{z})$ is unsatisfiable~\cite{DBLP:conf/tacas/ChristHN13,Christ_PhD}.
First, interpolants are derived for all theory lemmas: each lemma expresses that a conjunction of atoms from the original formula is unsatisfiable, these atoms can thus be divided into a conjunction $\alpha$ of atoms from $A$ and a conjunction $\beta$ of atoms from $B$, and an interpolant $I$ is derived for $\alpha \Rightarrow \neg \beta$.
Then, these interpolants are combined following the resolution proof of the solver.
This is how the interpolants from Ex.~\ref{ex:interpolation} were produced by the solvers.

The problem is therefore: given $A(\ve{x},\ve{y}) \land B(\ve{y},\ve{z})$ unsatisfiable, where $A$ and $B$ are conjunctions, how do we find $I(\ve{y})$ such that $A \Rightarrow I$ and $I \Rightarrow \neg B$? If the theory is linear rational arithmetic, this amounts to finding a separating hyperplane between the polyhedra $A$ and $B$. Let us note
\begin{equation}
\begin{array}{l@{\qquad}l}
A \defn \bigwedge_i \dotp{\ve{a''}_i}{\ve{x}} + \dotp{\ve{a}_i}{\ve{y}} \geq a'_i \,, &
B \defn \bigwedge_j \dotp{\ve{b''}_j}{\ve{z}} + \dotp{\ve{b}_j}{\ve{y}} \geq b'_j \,.
\end{array}
\end{equation}
Each $a'_i$ (resp. $b'_j$) is a pair $({a'}_i^\RR,{a'}_i^{\epsilon})$ lexicographically ordered, where ${a'}_i^\RR$ is the real part and ${c'}_i^{\epsilon}$ is infinitesimal; all other numbers are assumed to be real. $y \geq (x^{\RR},x^{\epsilon})$  with $x^{\epsilon} > 0$ and $y \in \RR$ expresses that $y > x^{\RR}$.

Since $A \land B$ is unsatisfiable, by Farkas' lemma, there exists an unsatisfiability witness $(\lambda_i)$, $(\mu_j)$, such that
\begin{equation}
\begin{array}{rl@{\qquad}rl}
\sum_i \lambda_i \ve{a}''_i & = 0 &
\sum_j \mu_j \ve{b}''_j & = 0\\
\sum_i \lambda_i \ve{a}_i + \sum_j \mu_j \ve{b}_j & = 0 &
\sum_i a'_i + \sum_j b'_j & > 0
\end{array}
\end{equation}
Such coefficients can in fact be read off the simplex tableau from the most common way of implementing a DPLL(T) solver for linear real arithmetic, as described in Sec.~\ref{sec:LRA}. Then the following is a valid interpolant (recall that the right-hand side can contain infinitesimals, leading to $>$):
\begin{equation}
I \defn \sum_i \dotp{\left(\lambda_i \ve{a}_i\right)}{\ve{y}} \geq
        \sum_i \lambda_i a'_i
\end{equation}

For polynomial arithmetic, one approach replaces nonnegative reals by sums-of-squares of polynomials, and Farkas' lemma by Positivstellensatz~\cite{DBLP:conf/cav/DaiXZ13}.

Another difficulty is posed for certain theories, for which the solving process involves generating lemmas introducing atoms not present in the original.
Consider the approaches for linear integer arithmetic described in Section~\ref{sec:LIA}:
except for branch-and-bound, all can generate new constraints involving any of the unknowns, without respecting the original partition of variables.
This poses a problem for interpolation: if interpolating for $A(x,y) \land B(y,z)$ over linear real arithmetic, we can rely on all atomic propositions being linear inequalities either over $x,y$ or $y,z$, but here, we have new atomic propositions that can involve both $x,z$.
Special theory-dependent methods are needed to get rid of these new propositions when processing the DPLL(T) proof into an interpolant~\cite{Christ_PhD,DBLP:conf/tacas/ChristHN13}.

\subsection{Optimization}
\label{sec:optimization}
Instead of finding one solution, one may wish to find a solution that maximizes (or nearly so) some function $f$.

A simple approach is \emph{binary search}: provided one can get a lower bound $l$ and an upper bound $h$ on the maximum $f(\ve{x}^*)$, one queries the solver for a solution $\ve{x}$ such that $f(\ve{x}) \geq m$, where $m = \frac{l+h}{2}$; if such a solution is found, refine the lower bound $l := f(\ve{x})$ and restart, otherwise $h := m$ and restart. Proceed until $l=h$.
This converges in finite time if $f$ has integer value.
This approach has been successfully applied to e.g. worst-case execution time problems~\cite{Henry_et_al_WCET_LCTES2014}.

In the case of LRA (resp. LIA), optimization generalizes \emph{linear programming} (resp. \emph{linear integer programming}) to formulas with disjunctions.
In fact, linear programming can be applied locally to a polyhedron of solutions:
when a DPLL(T) solver finds a solution $\ve{x}$ of a formula $F$, it also finds a conjunction $C$ of atoms such that $C \Rightarrow F$; $C$ defines a polyhedron and one can optimize within it, until a local optimum $\ve{x}^l$. Then one adds the constraint $f(\ve{x}) > f(\ve{x}^l)$ and restart; the last $\ve{x}^l$ found is the optimum (one can also detected unboundedness).
This approach can never enumerate the same $C$ (or subsets thereof) twice and thus must terminate.
It may, however, scan an exponential number of useless $C$'s; it may be combined with binary search for best effect~\cite{Sebastiani_Tomasi_IJCAR12}.

\section{Conclusion}
Considerable progress has been made within the last 15 years on increasingly practical decision procedures for increasingly large classes of formulas, even though worst-case complexity is prohibitive, and sometimes even though the class is undecidable.%
\footnote{Worst-case complexity, or completeness in complexity classes, is therefore not always a good indicator of practical performance. Average complexity is difficult to define (one needs to suppose a probability distribution on formulas) and may ill-describe practical use cases: it is well-known that random SAT instances behave unlike industrial examples \cite{Ach09HBSAT}, and same with random linear constraints~\cite{Monniaux_CAV09}. For want of better indication, performance is measured on libraries of benchmarks.}
Major ingredients to that success were
\begin{inparaenum}[i)]
\item lazy generation of lemmas, partial projections or instantiations, guided by counterexamples (as opposed to eager exhaustive generation, often explosive)
\item generalization of counterexamples so as to \emph{learn} sufficiently general blocking lemmas
\item tight integration of propositional and theory-specific reasoning.
\end{inparaenum}

Nonlinear arithmetic reasoning (polynomials, or even transcendental functions) is still a very open question. Current approaches in SMT \cite{Jovanovic_de_Moura_IJCAR2012} are based on partial cylindrical algebraic decomposition~\cite{Collins_CAD}; possibly methods based on critical points \cite{Grigor'ev:1988:SSP:53372.53375,Basu_Pollack_Roy,DBLP:conf/issac/DinS03} could be investigated as well.

There are several challenges to using computer algebra procedures inside a SMT solver.
\begin{inparaenum}[i)]
\item These procedures may not admit addition or retraction of constraints without recomputation.
\item They may compute eagerly large sets of formulas (as in conventional cylindrical algebraic decomposition).
\item They may be very complex and thus likely to contain bugs.%
\footnote{The author had several computer algebra packages crash or produce wrong results. Perhaps running large libraries of benchmarks would help in finding such bugs.}
Being able to produce independently-checkable proof witnesses would help in this respect.
\end{inparaenum}

\paragraph*{Acknowledgements} Thanks to the anonymous referees for their careful proofreading.

\printbibliography
\end{document}